\documentclass[final]{article}

     \usepackage[nonatbib]{neurips_2020}

\usepackage[utf8]{inputenc} 
\usepackage[T1]{fontenc}    
\usepackage[colorlinks=true,citecolor=blue]{hyperref}
\usepackage{url}            
\usepackage{booktabs}       
\usepackage{amsfonts}       
\usepackage{nicefrac}       
\usepackage{microtype}      
\usepackage{float}			
\usepackage{multirow}
\usepackage{graphicx}
\usepackage{wrapfig}
\usepackage{hyperref}
\usepackage{amsmath}
\usepackage{gensymb}
\usepackage{amssymb}

\usepackage[colorlinks=true,citecolor=blue]{hyperref}
\usepackage{color}

\newcommand{\real}{\mathbb{R}}
\newcommand{\x}{{\bf x}}
\newcommand{\y}{{\bf y}}
\newcommand{\z}{{\bf z}}

\newcommand{\remove}[1]{}

\newcommand{\mcalX}{\mathcal{X}}

\newcommand{\mcalH}{\mathcal{H}}

\newcommand{\bphi}{\boldsymbol{\mathit{\phi}}}

\newcommand{\bpsi}{\boldsymbol{\mathit{\psi}}}

\usepackage{graphicx}
\usepackage{dcolumn}
\usepackage{bm}

\usepackage{graphics}
\usepackage{pgf}
\usepackage{tikz}
\usetikzlibrary{arrows,automata,positioning}
\usetikzlibrary{mindmap,trees}

\usetikzlibrary{shapes,arrows,positioning} 

\tikzset{
    >=stealth',
    punkt/.style={
           rectangle,
           rounded corners,
           draw=black, very thick,
           text width=6.5em,
           minimum height=2em,
           text centered},
    pil/.style={
           ->,
           thick,
           shorten <=2pt,
           shorten >=2pt,}
}

\usepackage{array}
\newcolumntype{P}[1]{>{\centering\arraybackslash}p{#1}}

\title{Understanding Climate Impacts on Vegetation with  Gaussian Processes in Granger Causality}

\author{%
  Miguel Morata-Dolz, Diego Bueso, Maria Piles and Gustau Camps-Valls\thanks{\href{http://isp.uv.es}{http://isp.uv.es}} \\
  Image Processing Laboratory (IPL)\\
  Universitat de Val\`encia\\
  46980 Paterna (Val\`encia). Spain \\
  \texttt{\{miguel.morata,gustau.camps\}@uv.es}\\
}

\begin{document}

\maketitle

\begin{abstract}
Global warming is leading to unprecedented changes in our planet, with great societal, economical and environmental implications, especially with the growing demand of biofuels and food. Assessing the impact of climate on vegetation is of pressing need. We approached the attribution problem with a novel nonlinear Granger causal (GC) methodology and used a large data archive of remote sensing satellite products, environmental and climatic variables spatio-temporally gridded over more than 30 years.  
We generalize kernel Granger causality by considering the variables cross-relations explicitly in Hilbert spaces, and use the covariance in Gaussian processes. 
The method generalizes the linear and kernel GC methods, and comes with tighter bounds of performance based on Rademacher complexity. 
Spatially-explicit global Granger footprints of precipitation and soil moisture 
on vegetation greenness are identified more sharply than previous GC methods. 
\end{abstract}

\section{Introduction}

Establishing causal relations between random variables from observational data is perhaps the most important challenge in today's science in general and in Earth sciences in particular~\cite{Runge19natcom}. 
Granger causality (GC)~\cite{granger_investigating_1969} was introduced as a first attempt to formalize quantitatively the causal relation between time series, and is the most widely used method. The intuition behind GC is to test whether the past of $X$ helps in predicting the future of $Y$ from its past alone. GC implicitly tells us about the concept of {\em information} using {\em forecasting}. 
Other methods rely on similar concepts of {\em information flow} and {\em predictability}: connections can be established between GC and transfer entropy \cite{schreiber_measuring_2000}, 
convergent cross-mapping \cite{sugihara_detecting_2012}, 
and with the 
graphical causal model perspective \cite{white_linking_2011}. 
Noting the strong linearity assumption in GC~\cite{eichler_causal_2007}, nonlinear extensions of GC have been proposed. 
In particular, GC with kernels was originally introduced in \cite{Ancona}. The method assumed a particular class of functions and an additive interaction between them. An alternative kernel-based test in combination with a filtering approach 
was later introduced in \cite{marinazzo}. 
In all these studies, the autoregressive (AR) models use kernel-based regression on the concatenation of the involved variables in input spaces. This approach, however, is limited as it disregards nonlinear cross-relations between $X$ and $Y$ in Hilbert spaces explicitly.  
We here introduce explicit feature maps and corresponding kernel functions that account for nonlinear cross-relations in kernel space \cite{martinez06}. 

\section{Crosskernel Gaussian processes for Granger Causality}

GC first builds univariate and bivariate AutoRegressive (AR) models: 
(1) $y_{t+1} =  {\bf a}_t^\intercal {\bf y}_t + \varepsilon_t^y$ and 
(2) $y_{t+1} = {\bf a}_t^\intercal {\bf y}_t + {\bf b}_t^\intercal {\bf x}_t + \varepsilon_t^{y|x}$, where $\y_t =[y_t,y_{t-1},\ldots,y_{t-P}]^\intercal$, $\x_t= [x_t,x_{t-1},\ldots,x_{t-Q}]^\intercal$, and ${\bf a}=[a_1,\ldots,a_P]^\intercal$ and ${\bf b}=[b_1,\ldots,b_Q]^\intercal$ are typically estimated by least squares. A GC test is defined as the ratio of model fitting errors: $\delta_{x \to y} = \log({\mathbb V}[\varepsilon_t^y]/{\mathbb V}[\varepsilon_t^{y|x}])$, where the residual errors are defined for the unrestricted $\varepsilon_t^y$ and restricted $\varepsilon_t^{y|x}$ cases separately, and ${\mathbb V}$ represents the variance operator.  
Conditional Granger causality traditionally considers incrementing the variable {${\bf y_t}$ by stacking the variables one conditions to ${\bf z}$, that is ${\bf y_t}':=[{\bf y_t},{\bf z_t}]$, and applying the same methodology. } 
The linear GC formulation can be generalized to the nonlinear case using elements of the theory of reproducing kernel Hilbert spaces (RKHS)~\cite{Scholkopf02}. Let us assume the existence of a Hilbert space $\mcalH$ equipped with an inner product 
where samples in $\mcalX$ are mapped into by means of a feature map $\bphi:\mcalX\to\mcalH, \x_i\mapsto \bphi(\x_i)$, $1\leq i\leq n$. The similarity between the elements in $\mcalH$ can be estimated using its associated dot product $\langle\cdot,\cdot\rangle_{\mcalH}$ via RKHS, $k:\mcalX\times\mcalX\to\mathbb{R}$, such that  $(\x,\x')$ $\mapsto$ $k(\x,\x')$. 

\noindent{\bf Stacked kernel.} 
The standard kernel GC (KGC) approach considers a kernel-based AR modeling~\cite{marinazzo,Ancona}. 
The method defines two feature maps $\boldsymbol{\phi}$ and $\boldsymbol{\psi}$ to a RKHS ${\mathcal H}$ endorsed with reproducing kernels $k$ and $\ell$, where ${\bf y}_t$ and the concatenation $\z_t=[\y_t,\x_t]\in\real^{P+Q}$ are mapped to, respectively. This leads to the kernel regression models (1) $y_{t+1} = {\bf a}_H^\intercal \bphi(\y_t) + \varepsilon_t^y$ and (2) $y_{t+1} = {\bf b}_H^\intercal \bpsi(\z_t) + \varepsilon_t^{y|x}$, where now ${\bf a}_H,{\bf b}_H\in\real^{H\times 1}$. By using the representer's theorems ${\bf a}_H =  \boldsymbol{\Phi}^\intercal \boldsymbol{\alpha}$ and ${\bf b}_H = \boldsymbol{\Psi}^\intercal \boldsymbol{\beta},$ where $\boldsymbol{\Phi},\boldsymbol{\Psi}\in\real^{n\times H}$, the AR models can be defined in terms of kernel functions only:
$y_{t+1} = \boldsymbol{\alpha}^\intercal{\bf k}_t + \varepsilon_t^y$, and $y_{t+1} = \boldsymbol{\beta}^\intercal\boldsymbol{\ell}_t + \varepsilon_t^{y|x}$, respectively, where ${\bf k}_t=[k(\y_1,\y_t),\ldots,k(\y_n,\y_t)]^\intercal$ and $\boldsymbol{\ell}_t=[\ell(\z_1,\z_t),\ldots,\ell(\z_n,\z_t)]^\intercal$ contain all evaluations of 
$k$ and $\ell$ at time $t$. 
Since data are mapped to the same Hilbert space ${\mathcal H}$, the same kernel function and parameters are used for both $k$ and $\ell$.

\noindent{\bf Summation kernel.} 
Alternatively {\em implicit} AR models can be defined in RKHS~\cite{Ancona}: $y_{t+1} = {\bf a}_H^\intercal \bphi(\y_t) + \varepsilon_t^y$, and $y_{t+1} = {\bf a}_H^\intercal \bphi(\y_t) +{\bf b}_H^\intercal \bpsi(\x_t) + \varepsilon_t^{y|x},$
which leads to the kernel AR models 
$y_{t+1} = \alpha^\intercal{\bf k}_t + \varepsilon_t^y$ and $y_{t+1} = \alpha^\intercal{\bf k}_t + \beta^\intercal\boldsymbol{\ell}_t + \varepsilon_t^{y|x},$
where 
now $\boldsymbol{\ell}_t:=[\ell(\x_1,\x_t),\ldots,\ell(\x_n,\x_t)]^\intercal$. 
The summation kernel is more appropriate when large time embeddings $P$ and $Q$ are needed to capture long-term memory processes, since it avoids constructing large dimensional feature vectors $\z$ by concatenation. 
However, the cross-information between $X$ and $Y$ is missing. 

\noindent{\bf Explicit cross-kernel.} In order to account for cross-correlations in Hilbert space, while alleviating the issue of large embeddings in conditional GC setups.  
we propose to explicitly define two feature maps: the standard individual map $\phi$ and the joint feature mapping $\psi$ for the second AR model: $y_{t+1} = {\bf a}_H^\intercal \boldsymbol{\phi}(\y_t) + \varepsilon_t^y$ and $y_{t+1} = {\bf b}_H^\intercal \boldsymbol{\psi}(\x_t,\y_t) + \varepsilon_t^{y|x}$, 
where the joint map is defined by construction as
$\widetilde{\boldsymbol{\psi}}(\x_t,\y_t) := [{\bf A}_1\boldsymbol{\varphi}(\y_t),{\bf A}_2\boldsymbol{\varphi}(\x_t),{\bf A}_3(\boldsymbol{\varphi}(\y_t)+\boldsymbol{\varphi}(\x_t))]^\intercal,$ 
where $\boldsymbol{\varphi}$ is a nonlinear feature map into an RKHS ${\mathcal H}$, and ${\bf A}_i$, $i=1,2,3$, are three linear transformations from ${\mathcal H}$ to ${\mathcal H}_i$. The induced joint kernel function readily becomes:
\begin{eqnarray}
\begin{array}{ll}
&\hspace{-0.6cm}n((\x_t,\y_t),(\x_t',\y_t')) =\widetilde{\boldsymbol{\psi}}(\x_t,\y_t)^\intercal \widetilde{\boldsymbol{\psi}}(\x_t',\y_t') \\
 &\hspace{1cm}=\boldsymbol{\varphi}(\y_t)^\intercal{\bf R}_1\boldsymbol{\varphi}(\y_t') + \boldsymbol{\varphi}(\x_t)^\intercal{\bf R}_2\boldsymbol{\varphi}(\x_t') +\boldsymbol{\varphi}(\y_t)^\intercal{\bf R}_3\boldsymbol{\varphi}(\x_t') + \boldsymbol{\varphi}(\x_t)^\intercal{\bf R}_3\boldsymbol{\varphi}(\y_t')\\
&\hspace{1cm}=\!n_1(\y_t,\y_t')\!+\!n_2(\x_t,\x_t')\!+\!n_3(\y_t,\x_t')\!+\!n_4(\x_t,\y_t'),
\end{array}
\end{eqnarray}

\noindent where ${\bf R}_1={\bf A}_1^\intercal {\bf A}_1+{\bf A}_3^\intercal {\bf A}_3$, 
${\bf R}_2={\bf A}_2^\intercal {\bf A}_2 + {\bf A}_3^\intercal {\bf A}_3$, 
and ${\bf R}_3={\bf A}_3^\intercal {\bf A}_3$. The new kernel function considers cross-terms relations between the time series through kernels $n_3$ and $n_4$. 
Besides, there is no need to explicitly use the same kernel function or parameters, and can be convenient to alleviate the problem of increased dimensionality in conditional GC settings as ours. 

We used the previous kernel/covariance in Gaussian Processes (GPs)~\cite{rasmussen06}. 
The GP modeling of XKGC assumes that the AR functions $\mathbf{f}_y$ and  $\mathbf{f}_{y|x}$ follow $n$-dimensional Gaussian distributions $\mathbf{f}_y,\sim\mathcal{N}({\bf 0},{\bf K})$ and $\mathbf{f}_{y|x}\sim\mathcal{N}({\bf 0},{\bf L})$, and covariances ${\bf K}$ and ${\bf L}$ (or respectively ${\bf N}$) of the distributions are determined by a kernel function. 
A direct sum of covariances is also a covariance so all kernel functions in the XKGC framework ($k$, $l$ and $n$) induce valid GPs. 
The XKGC allows to advantageously optimize hyperparameters 
by Type-II Maximum Likelihood using the marginal likelihood of the observations. 
The GP treatment also permits to define a test statistic based on the the evidence. We suggest the GC criterion for the GP versions as follows: 
$$\delta^{\text{GP}}_{x\to y} = \max_{\boldsymbol{\theta}_2}\log(p_2({\bf f}_y|X,Y)) - \max_{\boldsymbol{\theta}_1}\log(p_1({\bf f}_y|Y)),$$
which represents the difference between log-evidences of the two GP AR models, so we infer $X\to Y$ when the evidence of conditioned model is larger than the evidence of the unconditioned model.

The cross-kernel GC (XKGC) generalizes previous KGC methods and comes with statistical guarantees when used in GPs. 
Owing to the connection between GPs and KRR \cite{kanagawa2018gaussian}, our GP model function class $h$ implicitly uses the squared loss for $\widehat{R}(h)$. Let us use an squared exponential kernel, $k({\bf x},{\bf x})=1$, and let $\gamma\in[0,1]$ and $\beta\in[\gamma,1]$. The Rademacher complexity regression minimum bound for the cross-kernel is:
$$R_{\text{cross}}(h) \leq \widehat{R}(h) + \frac{8\|h\|^2}{\sqrt{n}}\left(\sqrt{\frac{1+\gamma}{1+\beta}}+\frac{3}{4}\sqrt{\frac{\log(2/\delta)}{2}}\right),$$
which follows from the Rademacher complexity for a sum of $N$ kernels $K_i$ can be easily bounded as $\widehat R(h) = \sqrt{N} \widehat R(h_i)$, $i=1,\ldots,N$; $M^2=2(1+\beta)$ and the \text{Tr}$[{\bf K}]=2n(1+\gamma)$ for the cross-kernel. For the stacked kernel, $M^2=1$ and \text{Tr}$[{\bf K}]=n$, and for the summation $M^2=2$ and \text{Tr}$[{\bf K}]=2n$, we obtain $R_{\text{cross}}(h)\leq R_{\text{sum}}(h) = R_{\text{stacked}}$. 
Note that for $\gamma=\beta$, i.e. when $X$ and $Y$ convey correlated information, the cross-kernel bound converges to the stacked and the summation bounds. Since $\gamma\leq \beta$, the cross-kernel bound will be always tighter than the stacked/summation kernel bound.

\section{Experimental results}

\subsection{Data collection and preprocessing}

Our study used the database in \cite{Papagiannopoulou2017} available at \url{https://sat-ex.ugent.be/}, which consists of climatic data obtained from both satellite observations and in-situ measurements. Variables are classified into four categories of driving forcings: temperature, precipitation, soil moisture, and radiation. A total of $18$ different products are available, and span over $1981-2010$ on a global scale. Data were converted to a common monthly temporal resolution and $1º\times 1º$ latitude-longitude of spatial resolution, and $13072$ (pixel) time series were processed, see Appendix. 

The application of Granger causality requires stationary data so we removed the trend and periodicity of all time series. Both are confounding factors that inflate Granger detection. The seasonal cycle was estimated as the average annual mean over the $30$ years, and subtracted from the raw time series to yield time series of anomalies.  
Variable selection was also very relevant. We selected one product per variable. For this, a correlation study was carried out between all products with the NDVI product. A maximum vote strategy was deployed on the results of Pearson's, Spearman's and Kendall's correlation coefficients. 
This yielded a database of four predictor variables (temperature, precipitation, soil moisture and radiation) corresponding to the anomalies of the four categories. We performed the study with data from coincident months. Data standardization was applied to all series. 
Granger causality was then carried out using one variable at a time and conditioning on the others. Our target variable is NDVI that accounts for vegetation greenness. We compare the detection ability and class-specificity of the standard Vector Autoregressive (VAR) model, 
Gaussian Process (GP) and the proposed cross-kernel Gaussian Process (XKGP). All $13072$ models were cross-validated.

\subsection{Detection, robustness and class-specificity}

\begin{wrapfigure}{r}{5cm}
\centering
\vspace{-0.7cm}
\includegraphics[width=5cm]{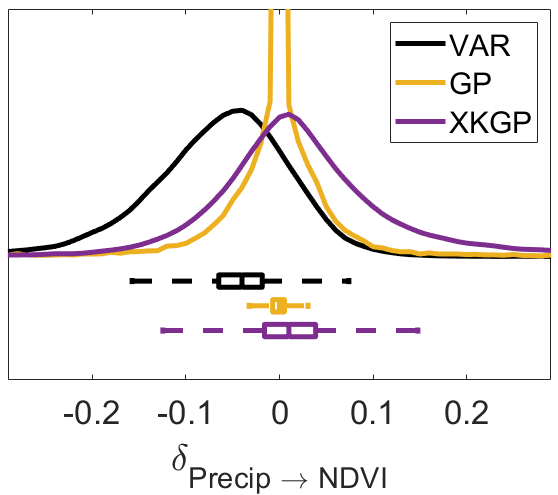}
\vspace{-0.6cm}
\caption{Density of global $\delta$ values between precipitation and NDVI. \label{fig:densities}}
\vspace{-0.5cm}
\end{wrapfigure}
Figure~\ref{fig:densities} shows the global densities of  $\delta$ values obtained for each regression model in the particular case of precipitation$\to$greenness (similar results were obtained for the case of soil moisture). The VAR model shows very poor detections globally (negative mean and heavy tail skewed over negative values) which indicates it is unable to find the causal relation in most of the pixels. The pathological case of standard GP indicates that the vast majority of decisions are slightly positive, yet close to zero, and the density is symmetric so detection is fairly compromised. The proposed XKGP leads to improved results over both VAR and GPs, with a clear positive mean and larger variability Granger-causal detections.

The results are analyzed per biome type for precipitation and soil moisture as driving forces in Table~\ref{tab:igbp}. The dominant capabilities of XKGP are obvious in both forcings (bold faced). Precipitation impacts greenness and XKGP finds stronger detections in shrublands, savannas and herbaceous (highlighted in italics), indicating enhanced detection capabilities in water-limited biomes. Similar results are observed when considering SM as the driver (conditioned to temperature, radiation and precipitation too). The lowest $\delta$ detections were obtained for forests, especially for needle-leaf and evergreen forests, as expected. 

\begin{table}[h!]
\centering
\small
\renewcommand{\arraystretch}{0.9}
\setlength{\tabcolsep}{4pt}
\caption{Mean $\delta$ and standard error ($\times$100) per IGBP land cover class and model for precip and SM.}
\label{tab:igbp}
\begin{tabular}{l|ccc|ccc}
\toprule
                           & \multicolumn{3}{c}{Precipitation}                    & \multicolumn{3}{|c}{Soil moisture}                     \\ \midrule
                           & VAR        & GP                 & XKGP               & VAR        & GP                 & XKGP  
                           \\
                           \midrule
Needleleaf Forest          & -6.2 $\pm$ 0.4 & 0.0 $\pm$ 0.1          & \textbf{0.3 $\pm$ 0.4}          & -6.2 $\pm$ 0.5 & 0.06 $\pm$ 0.04        & \textbf{0.9 $\pm$ 0.4}          \\
Evergreen Broadleaf Forest & -7.4 $\pm$ 0.2 & -0.1 $\pm$ 0.1         & \textbf{0.8 $\pm$ 0.3}          & -7.3 $\pm$ 0.3 & 0.3 $\pm$ 0.1          & \textbf{1.9 $\pm$ 0.3}          \\
Deciduous Broadleaf Forest & -7.0 $\pm$ 1.0     & -0.1 $\pm$ 0.1         & \textbf{0.7 $\pm$ 0.5}          & -9.0 $\pm$ 1.0     & -0.1 $\pm$ 0.2         & \textbf{2.0 $\pm$ 1.0}              \\
Mixed Forest               & -6.3 $\pm$ 0.4 & 0.0 $\pm$ 0.1          & \textbf{1.2 $\pm$ 0.4}          & -6.6 $\pm$ 0.4 & 0.1 $\pm$ 0.1          & \textbf{2.8 $\pm$ 0.4}          \\
Shrublands                 & \textit{-4.2 $\pm$ 0.3} & \textit{1.4 $\pm$ 0.3}          & \textit{\textbf{4.1 $\pm$ 0.7}}          & \textit{-2.7 $\pm$ 0.4} & \textit{2.4 $\pm$ 0.3} & \textit{\textbf{5.1 $\pm$ 0.6}} \\
Savannas                   & \textit{-3.6 $\pm$ 0.6} & \textit{0.8 $\pm$ 0.3}          & \textit{\textbf{5.9 $\pm$ 0.9}} & \textit{-5.3 $\pm$ 0.6} & \textit{0.6 $\pm$ 0.3}          & \textit{\textbf{5.9 $\pm$ 0.8}} \\
Herbaceous                 & \textit{-2.9 $\pm$ 0.7} & \textit{1.7 $\pm$ 0.6} & \textit{\textbf{6.8 $\pm$ 0.9}} & \textit{-3.9 $\pm$ 0.4} & \textit{1.0 $\pm$ 0.2}          & \textit{\textbf{7.1 $\pm$ 0.9}} \\
Cultivated                 & -5.4 $\pm$ 0.4 & 0.0 $\pm$ 0.2          & \textbf{3.2 $\pm$ 0.6}          & -4.3 $\pm$ 0.4 & 0.3 $\pm$ 0.1          & \textbf{6.8 $\pm$ 0.7}  \\ \bottomrule
\end{tabular}
\end{table}

\subsection{Global footprints of precipitation and moisture on vegetation}

\begin{figure}[H]
\centering
\includegraphics[width=0.9\textwidth]{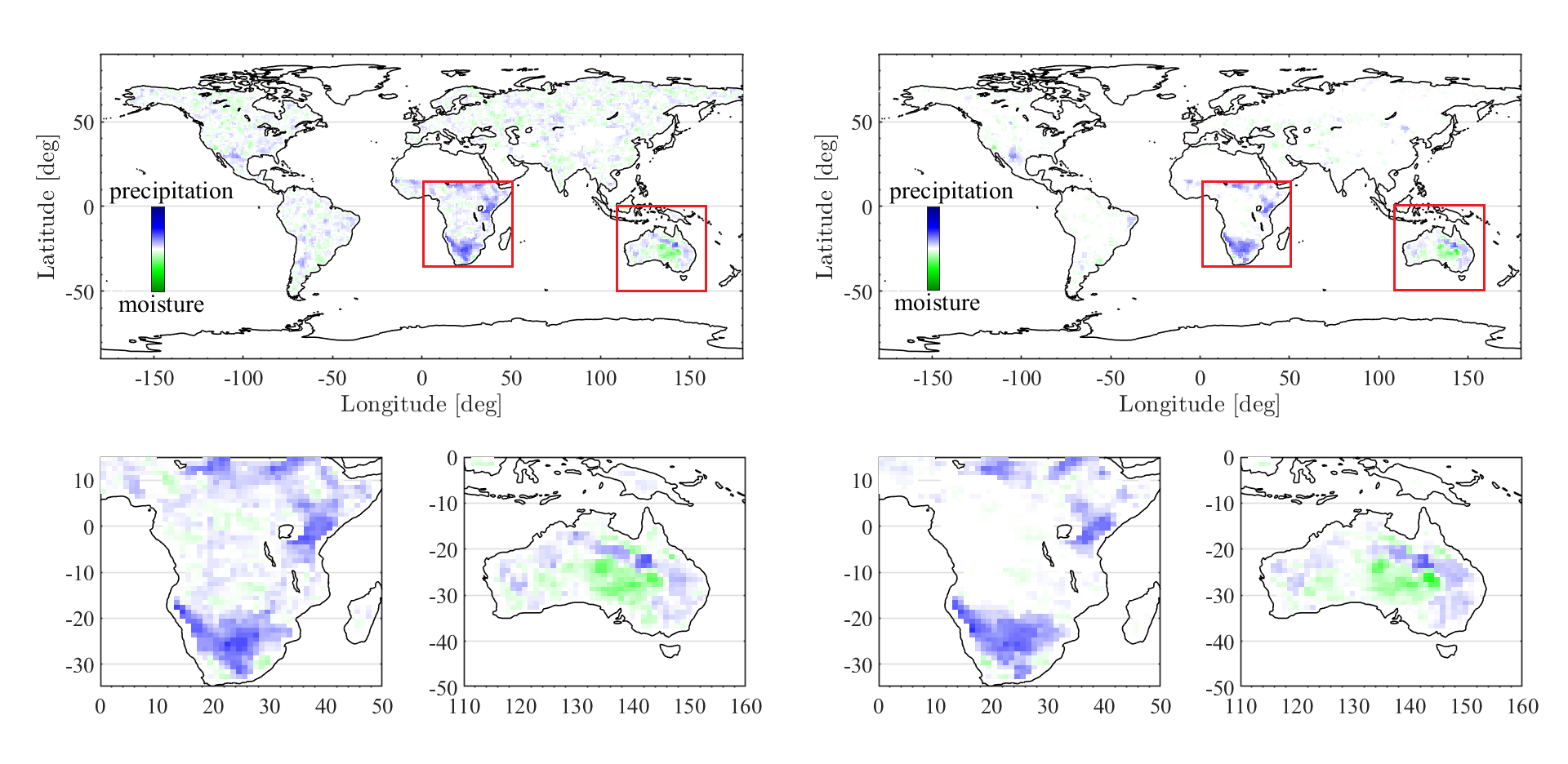}
\vspace{-0.6cm}
\caption{Global Granger footprint maps of precipitation and humidity on vegetation using VAR (left) and GP (right), with zoomed areas over Africa and Australia (bottom).\label{fig:global}}
\end{figure}

Let us now look at global and regional scales in Fig.~\ref{fig:global}. The gradient maps show where precipitation or moisture become more relevant for VAR (left) and GP (right). 
Overall, one can see many (spurious) detections of VAR, which cannot cope with the well-known nonlinear processes involved as reported in  \cite{Papagiannopoulou2017}. This can be better observed in the regional maps of Africa and Australia, where the GP model yields sharper and clearer detections spatially.
This suggests that the variability of central Africa, among other areas, obtained through VAR may be an artifact of the model. It is worth noting that areas detected with high GC impact correspond to those related to El Ni{\~n}o Southern Oscillation (ENSO) event, which causes droughts in the south-east of Africa and the east coast of Australia. 

Results show that Granger-climatic dynamics cause vegetation anomalies over most of the continental surface, with a greater impact in subtropical regions and mid-latitudes. Water availability is the main factor driving NDVI anomalies worldwide, finding a great Granger-causal relationship in semi-arid areas or in areas where 
a large part of vegetation dynamics responds to rainfall. 
In general, our findings highlight a strong dependence of global vegetation on water availability and show the effect of hydroclimatic anomalies on global vegetation over the studied period. These results suggest that vegetation 
is susceptible to follow future trends in water availability, and nonlinear (Granger) causal methods can capture this and quantify it adequately. 
These results suggest that vegetation will be critically influenced by the effect of climate warming on water-limited regions.
\vspace{-0.5cm}

\section{Conclusions}
We considered Granger causality and its wide adoption and applicability in Earth sciences. Noting the main shortcomings of stationarity and linearity, we proposed a kernel-based framework that generalizes previous linear and kernel GC approaches.  
The methodology copes with nonlinear relationships more efficiently and comes with statistical guarantees. The application on assessing the impact of climate variables on vegetation status and health summarized by vegetation indices suggested sharper and more robust Granger detection capabilities compared to previous methods.

\acksection
Miguel Morata-Dolz and Gustau Camps-Valls were supported by the European Research Council (ERC) under the ERC-2017-StG-SENTIFLEX (755617) and the ERC-CoG-2014 SEDAL (grant agreement 647423) projects, respectively. Authors thank the \href{https://sat-ex.ugent.be}{SAT-EX team} for granting access to the data used in this work.




\clearpage

\section*{Appendix. Data collection and availability\label{sec:data}}

\begin{table}[h!]
\centering
\caption{Data sets used in the analysis. These data sets are used to construct predictive features for the non-linear Granger causality framework. The NDVI is used to derive the target variable.}
\label{tab:my-table2}
\resizebox{\textwidth}{!}{%
\begin{tabular}{@{}lllll@{}}
\toprule
Variable                       & Dataset     & Spatial   resolution & Temporal   resolution & Reference                   \\ \midrule
\multirow{7}{*}{Temperature}   & CRU-HR      & 0.5º                 & monthly               & Harris et   al, 2014 \cite{Harris2014}        \\
                               & UDel        & 0.5º                 & monthly               & Willmott y Matsuura, 2001 \cite{Willmott2001}   \\
                               & ISCCP       & 1º                   & daily                 & Rossow y   Duenas, 2004 \cite{Rossow2004}     \\
                               & ERA-Interim & 0.75º                & 3- hourly             & Dee et al, 2011 \cite{Dee2011}            \\
                               & GISS        & 2º                   & monthly               & Hansen et   al, 2013 \cite{Hansen2013}       \\
                               & MLOST       & 5º                   & monthly               & Smith et al, 2008 \cite{Smith2008}          \\
                               & CFSR-Land   & 0.5º                 & daily                 & Coccia et   al, 2015 \cite{Coccia2015}       \\
\multirow{7}{*}{Precipitation} & CRU-HR      & 0.5º                 & monthly               & Harris et al, 2014 \cite{Harris2014}         \\
                               & UDel        & 0.5º                 & monthly               & Willmott y   Matsuura, 2001 \cite{Willmott2001} \\
                               & CPC-U       & 0.25º                & daily                 & Xie et al, 2007 \cite{Xie2007}            \\
                               & GPCC        & 0.5º                 & monthly               & Schneider   et al, 2016 \cite{Schneider2016}    \\
                               & CMAP        & 2.5º                 & monthly               & Xie y Arkin, 1997 \cite{Xie1997}          \\
                               & GPCP        & 2.5º                 & monthly               & Adler et   al, 2003 \cite{Adler2003}        \\
                               & MSWEP       & 0.25º                & 3- hourly             & Beck et al, 2017 \cite{Beck2017}           \\
Soil moisture                  & GLEAM       & 0.25º                & daily                 & Miralles   et al, 2011 \cite{Miralles2011}     \\
\multirow{2}{*}{Radiation}     & SRB         & 1º                   & 3- hourly             & Stackhouse et al, 2004 \cite{Stackhouse2004}     \\
                               & ERA-Interim & 0.75º                & 3- hourly             & Dee et al,   2011 \cite{Dee2011}          \\
\midrule
NDVI                           & GIMMS       & 0.25º                & monthly               & Tucker et al, 2005 \cite{Tucker2005}         \\ \bottomrule
\end{tabular}%
}
\end{table}

\end{document}